\documentclass[11pt]{article}

\usepackage{hyperref}
\usepackage{lipsum}
\usepackage[british]{babel}
\usepackage[utf8]{inputenc}
\usepackage{amsmath}
\usepackage{amssymb}
\usepackage{mathtools}
\usepackage{braket}
\usepackage{graphicx}
\usepackage{xcolor}
\usepackage{mathrsfs}
\usepackage{authblk}

\begin{document}
	
\title{A Schr\"odinger Equation for Light}
\author{Daniel R. E. Hodgson}
\affil{The School of Physics and Astronomy, University of Leeds, Leeds LS2 9JT, United Kingdom}

\maketitle

\begin{abstract}
In this chapter we examine the quantised electromagnetic (EM) field in the context of a Schrödinger equation for single photons. For clarity we consider only a one-dimensional system.  As a universal tool for calculating the time-evolution of quantum states, a Schrödinger equation must exist that describes the propagation of single photons.  Being inherently relativistic, however, critical aspects of both special relativity and quantum mechanics must be combined when quantising the EM field.  By taking the approach of a Schrödinger equation for localised photons, we will show how novel and previously overlooked features of the quantised EM field become a necessary part of a complete description of photon dynamics.  In this chapter, I shall provide a thorough examination of new features and discuss their significance in topics such as quantum relativity and photon localisation.  
\end{abstract}

\section{Introduction}

Thomas Young's double slit experiment gives a simple but clear demonstration that light is certainly a wave.  The appearance of an alternating pattern of dark fringes is evidence of the destructive superposition of waves passing through different slits onto the screen behind.  In this classical experiment, the pattern emerging on the screen is generated by the interference between oscillating electromagnetic (EM) waves that are predicted by Maxwell's theory of electromagnetism.  The modification from a classical to a quantum theory, however, reinterprets these waves as oscillations of the probabilistic wave function for a collection of photons, the indivisible particles of light.  In order to form a complete description of how photons evolve, it is important that we are able to define a wave function for each photon wave packet describing its oscillations through both space and time.

By initially postulating that photons are discrete and countable objects, and that each photon has an energy proportional to its frequency, it is possible to derive complete expressions for the electric and magnetic field observables up to an overall phase \cite{Bennett}.  More conventional quantisation methods, however, take the reverse approach.  Here the field observables, not the photons, are the main focus of the quantisation process, which are obtained by means of canonical quantisation.  See, for example, Ref.~\cite{Tann}.  By adopting a Hamiltonian procedure, correspondence with classical physics can be maintained by imposing canonical commutation relations.  Moreover, working directly with quantised fields may be viewed as more fundamental than working with particles, which are not covariant objects.  Providing a wave function for the excitations of the field observables, however, has proven exceptionally challenging. 

It was proven in 1948 by Newton and Wigner that no position-dependent wave function existed for the photon \cite{NW, Jordan}.  More specifically, subject to certain conditions,  there was no photon position operator with which to define a basis of localised eigenstates for the wave function.  Since this time, the localisation of single photons has been researched extensively \cite{Sch, Hof, Ros, Jau, Amr, Pik, Cha, Wig}, but there is as yet no unanimous agreement on whether localisation is possible.  In the work of Fleming \cite{Fle1, Fle2}, for example, some of the relativistic properties of the Newton-Wigner (NW) operator were clarified, but localisation of the photon was again shown to be impossible.  Only more recently, by considering the longitudinal components of spin, Hawton has been able to show that a photon position operator with commuting components conjugate to the momentum operator can be defined \cite{Haw1, Haw2, Haw3}. 

It is also unclear how the photon wave function ought to be interpreted.  Early wave functions, such as the Landau-Peierls wave function \cite{Lan, Coo1, Coo2}, were criticised for being non-locally related to the electric and magnetic field observables.  Such a relationship emerges due to a disparity between the units for a probabilistic wave function and the field observables.  Wave functions locally related to the field observables have been studied in both the first and second quantisation regimes \cite{Bia2, Bia3, Bia4, Sip, Smi}.  In some schemes, such as those of Knight \cite{Kni} and Licht \cite{Lic1, Lic2}, a photon is only localised if the electric and magnetic field expectation values are also localised.  In this case, however, a single photon cannot be localised if the field observables do not commute \cite{Bia1}.  To overcome problems of non-locality, many authors have introduced non-local inner products, which lead to the use of non-standard and non-hermitian models \cite{Sou, Haw4, Haw5, Gro}.

Localisation in quantum theory is also closely connected to causality.  A photon wave packet that is localised to one region, for instance, cannot reach another until a time has elapsed no less than the distance between these regions divided by the speed of light in a vacuum.  The theorems of Hegerfeldt \cite{Heg1} and Malament \cite{Mal}, however, show that non-zero correlations between the position of a wave packet can be generated at speeds exceeding the speed of light.  The question that this raises about causality has been a large topic of research \cite{Heg2, Heg3, Hal, Per, Bar}, with a particular interest in causality in the transmission of radiation between two two-level atoms \cite{Bis, Pow, Rub, Val}.  Whilst many insist that only causality in the sense of no-signalling, rather than of strict causality, is necessary, a wave function can only be usefully and properly interpreted if the speed at which it propagates never exceeds the speed of light. 

In this chapter we explore a recent quantisation of the free one-dimensional EM field in the position representation \cite{Hod}.  The main focus of this quantisation will be the construction of single-photon wave packets in a basis of localised photonic excitations in the Schr\"odinger picture.  We determine an equation of motion for these excitations which leads to a Schr\"odinger equation for the photon.  By focussing on dynamics, new parameters are introduced that were previously neglected or overlooked.  This provides us with a fuller description of the quantised EM field.  For completeness, expressions for the EM field observables shall be constructed, and a comparison with standard quantisations shall be given.

\section{The classical EM field in one dimension}

The equations of motion for light are Maxwell's equations.  The solutions to these equations provide us with the expected dynamics of the quantised particles of the EM field.  The purpose of this section is to review the appropriate equations of motion and their solutions in one dimension, and to determine an expression for the energy of the EM field.   

\subsection{The dynamics of the EM field}

\subsubsection{Maxwell's equations}
Light consists of two real, mutually propagating vector fields: the electric field and the magnetic field.  In one dimension, the electric and magnetic fields propagate along a single axis parametrised by a position coordinate $x$.  The electric and magnetic fields measured at a position $x$ at a time $t$ are denoted $\textbf{E}(x,t)$ and $\textbf{B}(x,t)$ respectively.  

Although $\textbf{E}(x,t)$ and $\textbf{B}(x,t)$ are parametrised by a position along the $x$-axis only, the fields are oriented, or polarised, in the plane orthogonal to the direction of propagation.  By specifying a right-handed Cartesian coordinate system $(x,y,z)$, the electric and magnetic fields have components in the $y$ and $z$ directions only.  The components of the fields oriented along the $y$ ($z$)-axis shall be referred to as horizontally (vertically) polarised. The polarisation of the field is specified by a discrete parameter $\lambda = \mathsf{H}, \mathsf{V}$.   

In a dielectric medium of constant permittivity $\varepsilon$ and permeability $\mu$, and by denoting $c = (\varepsilon\mu)^{-1/2}$, the horizontally and vertically polarised components of $\textbf{E}(x,t)$ and $\textbf{B}(x,t)$ satisfy the following simplified forms of Maxwell's equations:
\begin{align}
	\label{Maxwell1}
	\frac{\partial}{\partial x} \text{E}(x,t) &= \pm \frac{\partial}{ \partial t}\text{B}(x,t)\\
	c^2\frac{\partial}{\partial x}\text{B}(x,t) &= \pm \frac{\partial}{\partial t}\text{E}(x,t).\nonumber
\end{align}
In both lines of Eq.~(\ref{Maxwell1}) above, the electric and magnetic fields have alternate polarisations.  The positive (negative) sign applies when the electric and magnetic fields are vertically (horizontally) and horizontally (vertically) polarised respectively.

\subsubsection{The wave equation}

Maxwell's equations (\ref{Maxwell1}) couple together different components of the electric and magnetic field vectors.  By combining these equations, we can construct a second-order differential equation for each of the four field components independently.  These equations are
\begin{align}
	\label{wave1}
	\left[\frac{\partial^2}{\partial x} - \frac{1}{c^2}\frac{\partial^2}{\partial t}\right]\textbf{E}(x,t) &= 0\\
	\left[\frac{\partial^2}{\partial x} - \frac{1}{c^2}\frac{\partial^2}{\partial t}\right]\textbf{B}(x,t) &= 0.\nonumber
\end{align}
Here  we have four identical equations of motion, one for each of the four components of the EM field.  The solutions to Eq.~(\ref{wave1}) will be examined in Section \ref{Sec:Maxwellsolutions}.

\subsection{The energy and momentum of the EM field}

\subsubsection{The energy observable}

At each point in space and time, the electric and magnetic fields exert a force on any charged matter present at that point.  For this reason the EM field is able to do mechanical work on the charged matter, and must therefore store a certain amount of energy. Taking this into account, explicit expressions for the energy and momentum of light in a dielectric medium can be determined.  By considering the work done by the fields on a charge current density in a dielectric medium, one can show that the total energy along the $x$-axis is given by the expression
\begin{equation}
	\label{energy1}
	H_{\text{energy}}(t) = \int_{-\infty}^{\infty}\text{d}x\frac{A}{2}\left\{\varepsilon |\textbf{E}(x,t)|^2 + \frac{1}{\mu}|\textbf{B}(x,t)|^2\right\}.
\end{equation}
Here $A$ is the area occupied by the field in the $y$-$z$ plane.

\subsubsection{The Poynting vector}
The energy stored in the EM field in a particular region is carried in the direction of propagation in the form of the Poynting vector $\mathbf{S}$.  Since in one dimension light can only propagate along the $x$-axis, the only non-zero component of the Poynting vector is the $x$ component, which is given by the expression
\begin{equation}
	\label{Poynting1}
	S(x,t) = \frac{1}{\mu}\Big[\text{E}_\mathsf{H}(x,t)\text{B}_\mathsf{V}(x,t) - \text{E}_\mathsf{V}(x,t)\text{B}_\mathsf{H}(x,t)\Big].
\end{equation}
In the above the $\mathsf{H}$ and $\mathsf{V}$ subscripts refer to the horizontally and vertically polarised components of the fields respectively.  The expressions above, particularly Eq.~(\ref{energy1}), will be of importance in Section \ref{Sec:energyobservable}.

\subsection{The solutions to Maxwell's equations}
\label{Sec:Maxwellsolutions}

\subsubsection{Left- and right-propagating waves}

The wave equation (\ref{wave1}) describes the propagation of a wave along the $x$-axis at a constant speed $c$.  This is the speed of light in the medium.  In only one dimension, the solutions of the wave equation take a simple form.  By considering first the components of the electric field $\text{E}_{\lambda}(x,t)$, where $\lambda = \mathsf{H}, \mathsf{V}$, one can show that the expressions
\begin{equation}
	\label{solution1}
	\text{E}_{\lambda}(x,t) = \sum_{s=\pm1}\text{E}_{s\lambda}(x,t)
\end{equation}
satisfy Eq.~(\ref{wave1}) when $\text{E}_{s\lambda}(x,t) = \text{E}_{s\lambda}(x-sct,0)$.  In Eq.~(\ref{solution1}) above, the parameter $s =\pm1$ is introduced in order to differentiate between solutions propagating to the left (decreasing $x$) or the right (increasing $x$).  In this notation, light characterised by $s = -1 (+1)$ propagates to the left (right).  The exact form of $\text{E}_{s\lambda}(x,t)$ is determined from the initial conditions of the system. 

\subsubsection{Complete electric and magnetic field solutions} 

The corresponding magnetic field solution to Eq.~(\ref{wave1}) is not independent of the electric field solution.  By using Maxwell's equations (\ref{Maxwell1}), the magnetic field can be determined directly from the electric field solution (\ref{solution1}).  After taking into account the sign difference for different polarisations, one may show that
\begin{equation}
	\label{Efield1}
	\textbf{E}(x,t) = \sum_{s=\pm1}c\left[\text{E}_{s\mathsf{H}}(x,t)\widehat{\boldsymbol{y}} + \text{E}_{s\mathsf{V}}(x,t)\widehat{\boldsymbol{z}}\right]
\end{equation}
and 
\begin{equation}
	\label{Bfield1}
	\textbf{B}(x,t) = \sum_{s=\pm1}s\left[-\text{E}_{s\mathsf{V}}(x,t)\widehat{\boldsymbol{y}} + \text{E}_{s\mathsf{H}}(x,t)\widehat{\boldsymbol{z}}\right].
\end{equation}
Here $\widehat{\boldsymbol{y}}$ and $\widehat{\boldsymbol{z}}$ are unit vectors oriented in the positive $y$ and $z$ directions respectively.

\subsubsection{Energy and momentum}

Since $\textbf{E}(x,t)$ and $\textbf{B}(x,t)$ are both characterised by the solutions $\text{E}_{s\lambda}(x,t)$, the energy and Poynting vector of the field must also be characterised by these solutions.  Substituting Eqs.~(\ref{Efield1}) and (\ref{Bfield1}) into Eqs.~(\ref{energy1}) and (\ref{Poynting1}) one finds that 
\begin{equation}
	\label{energy2}
	H_{\text{energy}}(t) = \sum_{s=\pm1}\sum_{\lambda = \mathsf{H}, \mathsf{V}}\int_{-\infty}^{\infty}\text{d}x\;A\varepsilon c^2|\text{E}_{s\lambda}(x,t)|^2
\end{equation}
and
\begin{equation}
	\label{Poynting2}
	S(x,t) = \sum_{s=\pm1}\sum_{\lambda = \mathsf{H}, \mathsf{V}} \frac{sc}{\mu} |\text{E}_{s\lambda}(x,t)|^2.
\end{equation}    
It is clear from Eq.~(\ref{Poynting2}) that a positive Poynting vector indicates propagation to the right whereas a negative Poynting vector indicates propagation to the left.

\section{The wave function of the Photon}

The equations of motion for light studied in the previous section apply the first set of constraints to the particle behaviour of light in a homogeneous and isotropic dielectric medium.  For a correct and natural interpretation of the photon wave function, the probability distribution of particles represented by the wave function must evolve identically to the classical wave packets of an EM wave.  In this section we construct a Fock space of localised bosonic excitations that provide a basis for constructing single-photon wave packets.  By imposing a constraint on the dynamics of these excitations in 1+1-dimensional space-time, a Schr\"odinger equation is formulated for the photon.  

\subsection{The parameter space of single-photon wave packets}

\subsubsection{Unitary time evolution}

Consider the propagation of a photon wave packet through the dielectric medium along the $x$-axis.  At an initial time $t=0$, we may represent this wave packet in the Hilbert space by a state vector $\ket{\psi_1(0)}$.  After a time $t$ has passed, the photon wave packet is now found in the time-evolved state
\begin{equation}
\label{evolution1}
\ket{\psi_1(t)} = U(t,0)\ket{\psi_1(0)}
\end{equation}
where $U(t,0)$ is the unitary time-evolution operator from time $t=0$ to time $t$.  As we have determined that light must propagate along the $x$-axis at a speed $c$, the unitary operator $U(t,0)$ transports the left- and right-moving components of the wave packet to the left or the right by an exact distance $ct$.  

\subsubsection{A complete parameter space}
\label{Sec:parameterspace}

If we consider two single-photon wave packets that are entirely distinguishable from each other, then their corresponding state vectors must be orthogonal.  When we localise two photons to different points along the $x$-axis, they are distinguishable from each other. Localised photon states, therefore, are orthogonal to one another, and it is natural to characterise them by their position along the $x$-axis.  In the same way, photons with different polarisations are distinguishable and their state vectors orthogonal.  Photon states are therefore also characterised by a polarisation $\lambda$.  In addition to this, states describing propagation in opposite directions must be orthogonal to one another, and must be characterised by the discrete parameter $s = \pm 1$.

\begin{figure}
\centering
\includegraphics[width = 1.0\textwidth]{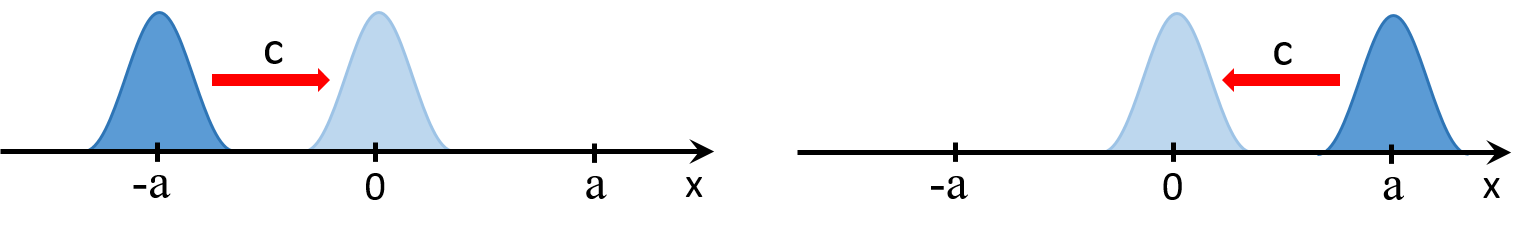}
\caption{The diagram illustrates the propagation of two localised wave packets.  In the left-hand diagram, a single photon propagates to the right from an initial position $x=-a$.  In the right-hand diagram, a single photon of identical shape propagates to the left from an initial position $x=a$.  At a later time both wave packets reach the origin.  Here the two wave packets are completely identical with respect to their position.}
\label{Fig:unitary}
\end{figure}

To see that this last parametrisation must be so, consider the setup illustrated in Figure \ref{Fig:unitary} showing two identical single-photon wave packets propagating in opposite directions.  We denote the state vectors for the left- and right-hand systems $\ket{\psi_1(x,t)}$ and $\ket{\psi_2(x,t)}$ respectively where $\ket{\psi_1(x,0)} = \ket{\psi_2(x+2a,0)}$.  At an initial time $t=0$, the photon in the left-hand diagram is localised to a position $x = -a$ whereas the photon in the right-hand diagram is localised to the position $x=a$.  Since the two wave packets occupy separate regions of the $x$-axis, their state vectors must be orthogonal:
\begin{equation}
\label{product1}
\braket{\psi_1(x,0)|\psi_2(x,0)} = \braket{\psi_1(x,0)|\psi_1(x-2a,0)} = 0.
\end{equation}

At a later time $t = a/c$, both photons will have travelled a distance $a$ to the left or the right of their initial positions.  After taking into account the direction of propagation of each of the photons, at this later time $t=a/c$ both wave packets will coincide with each other perfectly at the origin.  When parametrised by only a position and polarisation, at the time $t$ their corresponding state vectors will no longer be orthogonal. Hence, the inner product
\begin{equation}
\label{product2}
\braket{\psi_1(x,t)|\psi_2(x,t)} = \braket{\psi_1(x,t)|\psi_1(x-2a,t)}
\end{equation}
will necessarily be non-zero.  Given that the states $\ket{\psi_1(x,t)}$ and $\ket{\psi_2(x,t)}$ evolve unitarily according to Eq.~(\ref{evolution1}), however, the inner product between the two states at a time $t$ is given by 
\begin{align}
\braket{\psi_1(x,t)|\psi_2(x,t)} &= \braket{\psi_1(x,0)|U^\dagger(t,0)U(t,0)|\psi_2(x,0)}\\
&= \braket{\psi_1(x,0)|\psi_2(x,0)}\nonumber
\end{align}
where $\dagger$ denotes hermitian conjugation.  This inner product must be constant with respect to time.  The assumption that our two state vectors were initially orthogonal, therefore, is inconsistent with unitary time evolution and we reach a contradiction.

The resolution to this problem is to ensure that wave packets propagating in different directions remain orthogonal at all times.  To properly differentiate between states propagating in different directions, therefore, single-photon wave packets must also be parametrised by $s=\pm1$ in addition to position and polarisation.

\subsection{Local photons}

\subsubsection{Creation and annihilation operators}

During the interactions between light and matter, atoms absorb and emit light on the level of single photons.  The appropriate Hilbert space for the free EM field, therefore, is a Fock space of identical and non-interacting bosonic particles.  From what we have determined in the previous section, the localised photonic excitations of the EM field are characterised at any one time by a coordinate $x \in (-\infty,\infty)$, a polarisation $\lambda = \mathsf{H}, \mathsf{V}$ and a direction of propagation $s=\pm1$.  From now on we shall refer to such excitations as blips, which is the acronym for bosons localised in position.  

As is usual for a system of identical particles we may define a collection of blip annihilation operators that remove a single blip from the system.  The blip annihilation operator is denoted $a_{s\lambda}(x,t)$ in the Heisenberg picture and $a_{s\lambda}(x,0)$ in the Schr\"odinger picture.  A state containing only a single blip is defined
\begin{equation}
\label{blip1}
\ket{1_{s\lambda}(x,t)} = a^\dagger_{s\lambda}(x,t)\ket{0}.
\end{equation}
The operator $a^\dagger_{s\lambda}(x,t)$ is termed the blip creation operator and generates a single blip characterised by the parameters $(x,t,\lambda,s)$.  In Eq.~(\ref{blip1}) above, $\ket{0}$ is the vacuum state containing precisely zero blips.  The vacuum state satisfies the property
\begin{equation}
\label{vacuum1}
a_{s\lambda}(x,t)\ket{0} = 0
\end{equation}
for all $x$, $t$, $s$ and $\lambda$.

\subsubsection{Commutation relations}

Although the state defined in Eq.~(\ref{blip1}) contains only one blip, states containing an arbitrary number of blips can be generated by repeatedly applying the blip creation operators to the vacuum state.  Since blips are bosons, the resulting state must be unchanged through any reordering of the blips' positions.  Consequently, the ordering of these creation operators must be insignificant and they must commute with one another.  Hence
\begin{equation}
\label{comm1}
\Big[a^\dagger_{s\lambda}(x,t), a^\dagger_{s'\lambda'}(x',t')\Big] = 0 = \Big[a_{s\lambda}(x,t), a_{s'\lambda'}(x',t')\Big]
\end{equation}
for any $x$, $x'$, $t$, $t'$, $\lambda$, $\lambda'$, $s$ and $s'$.

In Section \ref{Sec:parameterspace} it was discussed how blips located at different positions, carrying different polarisations or propagating in opposite directions must be perfectly distinguishable from one another.  As a consequence, the states that represent them must also be orthogonal.  Taking this into account, we specify the following inner product for two single-blip states:
\begin{equation}
\label{comm2}
\braket{1_{s\lambda}(x,t)|1_{s'\lambda'}(x',t)} = \delta_{s,s'}\,\delta_{\lambda,\lambda'}\,\delta(x-x').
\end{equation}
Using Eqs.~(\ref{blip1}) and (\ref{vacuum1}), and expressing the inner product (\ref{comm2}) in terms of blip creation and annihilation operators, it can be shown that at any fixed time $t$
\begin{equation}
\label{comm3}
\left[a_{s\lambda}(x,t),a^\dagger_{s'\lambda'}(x',t)\right] = \delta_{s,s'}\,\delta_{\lambda,\lambda'}\,\delta(x-x').
\end{equation}
This is the fundamental commutation relation for blips. 

\subsubsection{The photon wave function}

In the context of linear optics experiments \cite{Zuk, Lim}, it is usual to talk about single photons when referring to particles whose state vectors $\ket{1(t)}$ can be expressed $\ket{1(t)} = a^\dagger(t)\ket{0}$ where $a(t)$ is an annihilation operator satisfying the commutation relation
\begin{equation}
\label{comm4}
\left[a(t),a(t)^\dagger\right] = 1.
\end{equation}
The blip states defined in Eq.~(\ref{blip1}) are not normalisable, but when superposed over a region of the $x$-axis can provide a localised basis for normalised single-photon wave packets.  Taking this into account, the annihilation operator for a single-photon wave packet can be defined in the following way: 
\begin{equation}
\label{wavefunction1}
a(t) = \sum_{s=\pm1}\sum_{\lambda = \mathsf{H}, \mathsf{V}}\int_{-\infty}^{\infty}\text{d}x\;\psi^*_{s\lambda}(x)\,a_{s\lambda}(x,t).
\end{equation}
Here $*$ denotes complex conjugation.  In Eq.~(\ref{wavefunction1}), the operator $a(t)$ is properly normalised and satisfies Eq.~(\ref{comm4}) when  \begin{equation}
\sum_{s=\pm1}\sum_{\lambda = \mathsf{H}, \mathsf{V}}\int_{-\infty}^{\infty}\text{d}x\;|\psi_{s\lambda}(x)|^2 = 1.
\end{equation}

The function $\psi_{s\lambda}(x)$ in Eq.~(\ref{wavefunction1}) represents the probability amplitude for finding a photon with polarisation $\lambda$ propagating in the $s$ direction at a position $x$.  More specifically, the transition probability between the single-photon state $\ket{1(t)}$ and the state $\ket{1_{s\lambda}(x,t)}$ is given by the expression
\begin{equation}
\label{wavefunction2}
\left|\,\braket{0|a_{s\lambda}(x,t)a^\dagger(t)|0}\,\right|^2 = |\psi_{s\lambda}(x)|^2.
\end{equation}
Hence $\psi_{s\lambda}(x)$ has the correct properties to be interpreted as a single-photon wave function in the position representation.

\subsection{A Schr\"odinger equation for light}

\subsubsection{A Hamiltonian constraint}

In order to calculate the dynamics of a quantum system, it is usual to first determine the Hamiltonian for that system.  In a closed system, the Hamiltonian would be given by the energy observable.  Once found, the Hamiltonian is used to construct a Schr\"odinger equation for state vectors in the Hilbert space.  So far, an energy observable has not been constructed for the blip states.  Moreover, there is no immediate choice for this observable, as, having complete uncertainty in their frequency, blips are not the eigenstates of the energy observable.  Fortunately, however, the dynamics of single blips have already been determined.  They are given by the solutions to Maxwell's equations (\ref{solution1}).

Blips are characterised by both a coordinate $x$ in space and a coordinate $t$ in time.  A single blip, therefore, may exist at one position at one moment in time, and then at a different position at another moment in time.  The classical dynamics of light in the medium places a constraint on which positions the blip may take from one moment to the next.  Being more specific, in order to satisfy Maxwell's equations, the expectation value of a localised blip must propagate at a speed $c$ along the $x$-axis without any dispersion.  These dynamics are imposed by the constraint $\braket{a_{s\lambda}(x,t)} = \braket{a_{s\lambda}(x-sct,0)}$.  Since this applies for any time-independent state, we can determine the general constraint
\begin{equation}
\label{dyn1}
a_{s\lambda}(x,t) = a_{s\lambda}(x-sct,0).
\end{equation}
When allowed to propagate freely, a blip found at $x$ at a time $t$ will be found at a position $x-sct$ at the time $t=0$.

The constraint on the dynamics, Eq.~(\ref{dyn1}), enables us to define an equation of motion for the blip operators $a_{s\lambda}(x,t)$.  More specifically, by taking the time derivative of Eq.~(\ref{dyn1}) it can be shown that
\begin{equation}
\label{dyn2}
\left[\frac{\partial}{\partial t}+sc\frac{\partial}{\partial x}\right]a_{s\lambda}(x,t) = 0.
\end{equation}
The equation above takes the form of a Wheeler-deWitt equation, and defines a stationary or ``timeless" state of the system \cite{Alt}.  In the system considered here, this equation confines the trajectories of blips to the boundaries of the light cone.  By relating a change in time to a change in the position of a blip in this way, we obtain a Schr\"odinger equation for blips:    
\begin{equation}
\label{Sch1}
i\hbar\frac{\partial}{\partial t}\ket{1_{s\lambda}(x,t)} = -i\hbar sc\frac{\partial}{\partial x}\ket{1_{s\lambda}(x,t)}.
\end{equation}

\subsubsection{The dynamical Hamiltonian}

In the context of a Schr\"odinger equation, the motion of the blip given by the right-hand side of Eq.~(\ref{Sch1}) is generated by the Hamiltonian for this system.  It is very convenient to determine this Hamiltonian as it provides a basis for introducing interactions in more complex models.  To this end, by using the Schr\"odinger equation for blips (\ref{Sch1}), we can determine exactly the Hamiltonian for the free propagation of light in a one-dimensional dielectric medium.  Here we shall denote this operator $H_{\text{dyn}}(t)$ with the subscript ``dynamical" to distinguish it as the Hamiltonian operator present in the Schr\"odinger equation.

Looking again at the right-hand side of Eq.~(\ref{Sch1}), it can be seen that the number of blips, their polarisation and their direction of propagation are all preserved as they evolve in time.  It is only their position that changes.  Taking this into account, a suitable ansatz for the dynamical Hamiltonian $H_\text{dyn}(t)$ would be
\begin{equation}
\label{dyn3}
H_{\text{dyn}}(t) = \sum_{s=\pm1}\sum_{\lambda = \mathsf{H}, \mathsf{V}}\int_{-\infty}^{\infty}\text{d}x\int_{-\infty}^{\infty}\text{d}x'\;i\hbar sc\,
f_{s\lambda}(x,x')\,a^\dagger_{s\lambda}(x,t)a_{s\lambda}(x',t)
\end{equation}
where $f_{s\lambda}(x,x')$ is a function to be determined.  This operator takes the form of an exchange operator that annihilates a blip at one position and replaces it with an identical blip at a different position.  To ensure that $H_\text{dyn}(t)$ is hermitian, $f_{s\lambda}(x,x') = -f_{s\lambda}(x',x)$.

In the Heisenberg picture, the dynamics of a blip operator $a_{s\lambda}(x,t)$ can be equivalently expressed through Heisenberg's equation of motion:
\begin{equation}
\label{Heis1}
\frac{\partial}{\partial t}a_{s\lambda}(x,t) = -\frac{i}{\hbar}\Big[a_{s\lambda}(x,t), H_{\text{dyn}}(t)\Big].
\end{equation}
Hence, by substituting the Hamiltonian (\ref{dyn3}) into Heisenberg's equation (\ref{Heis1}), making use of the commutation relations (\ref{comm1}) and (\ref{comm3}), and ensuring equivalence to Eq.~(\ref{dyn2}), it can be shown that
\begin{equation}
\label{dyn4}
f_{s\lambda}(x,x') = -\frac{\partial}{\partial x}\delta(x-x')
\end{equation}
and therefore
\begin{equation}
\label{dyn5}
H_{\text{dyn}}(t) = -i\sum_{s=\pm1}\sum_{\lambda = \mathsf{H}, \mathsf{V}}\int_{-\infty}^{\infty}\text{d}x\;\hbar sc\, a^\dagger_{s\lambda}(x,t)\frac{\partial}{\partial x}a_{s\lambda}(x,t).
\end{equation}
This Hamiltonian is hermitian and therefore a generator of unitary dynamics.  It should also be noted that the Hamiltonian for right-propagating blips takes the negative value of the Hamiltonian for left-propagating blips. This demonstrates that a right-propagating blip behaves identically to a left-propagating blip when the direction of time is reversed.

\section{Field observables in the position representation}

The approach to quantisation taken here differs from usual procedures by focussing on the particle character of the EM field rather than the quantised field observables.  It is by taking this point of view that the field observables do not require a direct relationship to the wave function of the photon.  This view is also held in Ref.~\cite{Fle3}.  The field observables remain, however, the fundamental observables from which we may derive expressions for the energy and momentum of the EM field.  The purpose of this section is to construct the electric and magnetic field observables in the position representation acting on the extended blip Hilbert space.  By insisting that blips are the localised excitations of the EM field, the field observables obtain unique characteristics that are crucial for a fuller understanding of many quantum effects.  

\subsection{The EM Field observables}

\subsubsection{An ansatz for the EM field observables}

The electric and magnetic field observables $\textbf{E}(x,t)$ and $\textbf{B}(x,t)$ are a linear and hermitian superposition of the creation and annihilation operators for the photonic excitations of the system.  In the position representation, these are the blip operators $a^\dagger_{s\lambda}(x,t)$ and $a_{s\lambda}(x,t)$.  Although this superposition is linear, there is no reason to assume that this superposition must be local.  In other words, the field observables at a position $x$ do not need to be a superposition of blip operators defined at that same point only.  For this reason, it is useful to introduce the notation
\begin{equation}
\label{reg1}
R_{s\lambda}(x,t) = \int_{-\infty}^{\infty}\text{d}x'\;\mathcal{R}_{s\lambda}(x,x')\,a_{s\lambda}(x',t).
\end{equation} 
Here $R_{s\lambda}(x,t)$ is referred to as the regularised annihilation operator and $\mathcal{R}_{s\lambda}(x,x')$ is a distribution over the $x$ axis.

In the following, the operators $\boldsymbol{\mathcal{E}}(x,t)$ and $\boldsymbol{\mathcal{B}}(x,t)$ shall denote the complex part of the electric and magnetic field observables respectively.  The total real fields are given by the hermitian superposition $\textbf{O}(x,t) = (\boldsymbol{\mathcal{O}}+ \boldsymbol{\mathcal{O}}^\dagger)/2$ where $\textbf{O} = \textbf{E}, \textbf{B}$ and $\boldsymbol{\mathcal{O}} = \boldsymbol{\mathcal{E}}, \boldsymbol{\mathcal{B}}$.  Taking this into account, an appropriate ansatz for the complex field observables is
\begin{equation}
\label{Efield2}
\boldsymbol{\mathcal{E}}(x,t) = \sum_{s=\pm1} c\Big[R_{s\mathsf{H}}(x,t)\widehat{\boldsymbol{y}} + R_{s\mathsf{V}}(x,t)\widehat{\boldsymbol{z}}\Big]
\end{equation}
and
\begin{equation}
\label{Bfield2}
\boldsymbol{\mathcal{B}}(x,t) = \sum_{s=\pm1} s\Big[-R_{s\mathsf{V}}(x,t)\widehat{\boldsymbol{y}} + R_{s\mathsf{H}}(x,t)\widehat{\boldsymbol{z}}\Big].
\end{equation}
It may be noted here that all components of the real electric and magnetic field observables commute.

\subsubsection{The regularisation function}

The regularisation function $\mathcal{R}_{s\lambda}(x,x')$ provides a relationship between the field observables and the blip operators.  Whilst in many quantisations photon wave packets must be locally related to the field observables, for a general choice of $\mathcal{R}_{s\lambda}(x,x')$, blips at one position may contribute to the field observables at another position.  In fact, we shall see later in this chapter that a single blip contributes to the field observables at all positions along the $x$-axis.  Notwithstanding this, the function $\mathcal{R}_{s\lambda}(x,x')$ must satisfy several general conditions.

Like the blip operators, the expectation values of the field observables must satisfy Maxwell's equations with respect to any time-independent state.  Taking into account the orientation of the field components in Eqs.~(\ref{Efield2}) and (\ref{Bfield2}), this condition implies that the regularised blip operators $R_{s\lambda}(x,t)$ must satisfy Eq.~(\ref{dyn2}).  Since this equation is also satisfied by the blip operators, using Eq.~(\ref{reg1}) it can be demonstrated that $\mathcal{R}_{s\lambda}(x,x')$ must be position invariant; that is, $\mathcal{R}_{s\lambda}(x,x') = \mathcal{R}_{s\lambda}(x-x')$.  What is more, since the medium is homogeneous and isotropic, the regularisation function must be symmetric, $\mathcal{R}_{s\lambda}(x-x') = \mathcal{R}_{s\lambda}(x'-x)$, and independent of $s$ and $\lambda$, $\mathcal{R}_{s\lambda}(x-x') = \mathcal{R}(x-x')$.  

\subsection{Energy in the position representation}
\subsubsection{The energy observable}

Now that we have a pair of expressions for the electric and magnetic field observables, it is possible to determine the energy observable for the free field in one-dimension.  To do so we substitute the field observables (\ref{Efield2}) and (\ref{Bfield2}) into the classical expression for the energy determined in Eq.~(\ref{energy1}).  We find that
\begin{equation}
\label{energy3}
H_{\text{energy}}(t) = \sum_{s=\pm1}\sum_{\lambda = \mathsf{H}, \mathsf{V}}\int_{-\infty}^{\infty}\text{d}x\;\frac{A\varepsilon c^2}{4}\left\{R_{s\lambda}(x,t) + H.c\right\}^2.
\end{equation}
Due to the square in the integrand, this observable is strictly positive as would be expected for an energy.  This result, however, implies that the energy observable cannot be equal to the dynamical Hamiltonian. Whereas the energy of a single blip is always positive, the left- and right-moving components of the dynamical Hamiltonian have opposite signs.  

\subsubsection{Energy conservation}
\label{Sec:energyobservable}

In a closed system, energy is always conserved.  In standard quantisations, when the dynamical Hamiltonian is equivalent to the energy observable, conservation of energy is guaranteed automatically as a consequence of Heisenberg's equation.  We have seen, however, that in this quantisation the dynamical Hamiltonian and the energy observable are not equal.  Energy conservation is only guaranteed, therefore, if $H_{\text{energy}}(t)$ and $H_{\text{dyn}}(t)$ commute.  Using the expressions for $H_\text{dyn}(t)$ and $H_{\text{energy}}(t)$ given in Eqs.~(\ref{dyn3}) and (\ref{energy3}) respectively, and by taking into account that $f_{s\lambda}(x-x')$ is an odd function and $\mathcal{R}(x-x')$ an even function, it can be shown that the dynamical Hamiltonian and the energy observable commute with each other.  Hence, the energy of the free EM field is conserved. 

\subsection{Non-local contributions to the field observables}

\subsubsection{Monochromatic excitations}

It can be seen from Eq.~(\ref{energy3}) that the regularisation operator plays an important role in determining the energy of a photon. Since the energy of a photon is determined by its frequency, it is convenient to express the energy observable (\ref{energy3}) in a basis of monochromatic excitations.  Such a set of excitations can be constructed by considering the Fourier transform of the localised blip operators.  We introduce, therefore, the operators
\begin{equation}
\label{Fourier1}
a_{s\lambda}(k,t) = \int_{-\infty}^{\infty}\frac{\text{d}k}{\sqrt{2\pi}}\;e^{-ikx}a_{s\lambda}(x,t)
\end{equation}
which, using Eq.~(\ref{comm3}), can be shown to satisfy the equal-time commutation relation
\begin{equation}
\label{comm5}
\left[a_{s\lambda}(k,t), a^\dagger_{s'\lambda'}(k',t)\right] = \delta_{ss'}\,\delta_{\lambda\lambda'}\,\delta(k-k').
\end{equation}
All annihilation operators commute amongst themselves, as do the creation operators.

\subsubsection{The energy of a photon}

Expressed in terms of the monochromatic operators $a_{s\lambda}(k,t)$ and their hermitian conjugates, the energy observable (\ref{energy3}) takes the alternative form 
\begin{equation}
\label{energy4}
H_{\text{energy}}(t) = \sum_{s=\pm1}\sum_{\lambda = \mathsf{H}, \mathsf{V}}\int_{-\infty}^{\infty}\text{d}k\;\frac{A\varepsilon \pi c^2}{2}\|\widetilde{\mathcal{R}}(k)\,a_{s\lambda}(k,t) + \widetilde{\mathcal{R}}^*(-k)\,a^\dagger_{s\lambda}(-k,t)\|^2.
\end{equation}
In the expression above, $\widetilde{\mathcal{R}}(k)$ is the Fourier transform of the regularisation function $\mathcal{R}(x-x')$ defined earlier. By taking into account the commutation relation (\ref{comm5}), we can show that the energy expectation value of a single monochromatic excitation of angular frequency $\omega = skc$ is
\begin{equation}
\label{energy5}
\braket{1_{s\lambda}(k,t)|H_{\text{energy}}(t)|1_{s\lambda}(k,t)} = A\varepsilon\pi c^2\,|\widetilde{\mathcal{R}}(k)|^2\,\delta(0).
\end{equation} 
In the above the state $\ket{1_{s\lambda}(k,t)}$ is defined analogously to the blip state in Eq.~(\ref{blip1}).  The delta function appearing in Eq.~(\ref{energy5}) is due only to the infinite normalisation of the monochromatic states.  

When an atom with transition energy $\hbar |\omega|$ emits a photon, exactly one excitation is generated oscillating with an angular frequency $\omega$.  If the total energy is carried away by the photon, an excitation of frequency $\omega$ must have an energy $\hbar |\omega|$.  Equating the energy of the monochromatic excitation, therefore, with the expectation value (\ref{energy5}), we can determine up to an overall phase an expression for the function $\widetilde{\mathcal{R}}(k)$.  Doing so we find that
\begin{equation}
\label{reg2}
\widetilde{\mathcal{R}}(k) = \sqrt{\frac{\hbar|k|}{A\varepsilon \pi c}} 
\end{equation}

\subsubsection{A non-local regularisation}

Now that we have determined $\widetilde{\mathcal{R}}(k)$ we can calculate $\mathcal{R}(x-x')$ explicitly, thus providing us with a relationship between blips and the field observables in the position representation.  Taking the Fourier transform of Eq.~(\ref{reg2}), we find that 
\begin{equation}
\label{reg3}
\mathcal{R}(x-x') = \int_{-\infty}^{\infty}\frac{\text{d}k}{2\pi}\;e^{ik(x-x')}\sqrt{\frac{2\hbar|k|}{\varepsilon A c}}.
\end{equation}
When $x\neq x'$, this expression can be given in the alternative form
\begin{equation}
\label{reg4}
\mathcal{R}(x-x') = -\sqrt{\frac{\hbar}{4\pi\varepsilon Ac}}\frac{1}{|x-x'|^{3/2}}.
\end{equation}
The distribution $\mathcal{R}(x-x')$ is non-vanishing everywhere, and decreases away from the origin with the negative three halves power of distance.

\begin{figure}
\centering
\includegraphics[width = 0.75\textwidth]{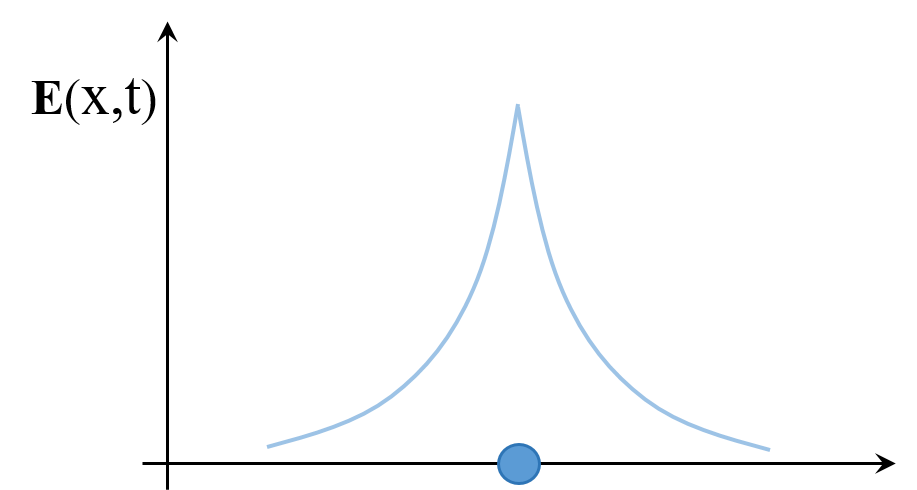}
\caption{The figure illustrates the contribution of a single blip (the blue spot) to the electric field observable.  The blip in the diagram contributes to measurements of the electric field observable at all points along the $x$ axis.  The magnitude of the contribution decreases with the distance of the measurement from the blip.}
\label{Fig:fields}
\end{figure}

Whilst the blips represent the localised particles of the system, and interact only locally with their surroundings, they contribute to the measurement of the electric and magnetic field observables in a highly non-local way.  One way to think about this relationship is that the measurement of the electric and magnetic fields takes into account the behaviour of blips at all positions in space.  An alternative perspective would be to treat each blip as a ``carrier" of a non-local field.  This interpretation is visualised in Figure \ref{Fig:fields} showing the contribution of a single blip to the electric field.  In a similar manner to as one may think about the gravitational field about the earth, one may think of a single blip as being surrounded by and carrying with it a non-local electromagnetic field.

\section{Conclusions}
	
In standard treatments of the quantised EM field in one dimension, the basic excitations of the field are characterised by a wave vector $k$ and a polarisation $\lambda$.  By demanding that photons can be localised, however, and that they must propagate at the speed of light along the $x$-axis, we have shown that an additional discrete parameter $s =\pm1$ must be introduced, thus doubling the size of the usual Hilbert space.  Whilst canonical quantisation overlooks these degrees of freedom, by taking the viewpoint of a Schr\"odinger equation, these additional degrees of freedom arise naturally.  What is more, by maintaining the perspective of a Schr\"odinger equation, a Hamiltonian operator $H_\text{dyn}(t)$ was constructed that generates the time-evolution of state vectors in the Hilbert space.  Unlike other quantisations, our Hamiltonian necessarily possesses both positive and negative eigenvalues.  This can be most clearly seen by expressing the dynamical Hamiltonian in the momentum representation.  In terms of the monochromatic operators defined in Eq.~(\ref{Fourier1}), $H_{\text{dyn}}(t)$ is given by the expression
\begin{equation}
\label{con1}
H_{\text{dyn}}(t) = \sum_{s=\pm1}\sum_{\lambda = \mathsf{H}, \mathsf{V}}\int_{-\infty}^{\infty}\text{d}k\;\hbar sck\,a^\dagger_{s\lambda}(k,t)a_{s\lambda}(k,t).  
\end{equation}
This Hamiltonian is almost identical to the usual Hamiltonian of the free EM field, but the eigenvalues $\hbar s ck$ can take both positive and negative values.
	
In standard quantisations, the Hamiltonian is always positive.  The Hamiltonian above is positive only when $s$ and the sign of $k$ coincide, or equivalently when the sign of $k$ indicates direction of propagation.  Under this condition the dynamical Hamiltonian (\ref{con1}) and the energy observable (\ref{energy4}) are equal.  In this quantisation we therefore go beyond current assumptions by demanding that the Hamiltonian can be both positive and negative.  This idea is not new, however. In the work of Hegerfeldt and others, it was shown that a wave packet cannot propagate causally when the Hamiltonian is bounded from below \cite{Heg3}.  Mostafazadeh and Zamani \cite{Mos}, therefore, introduced a new inner product that enabled the use of negative frequency states.  More recently, with similar justifications, Hawton considers real EM field excitations that necessarily contain both positive and negative frequency contributions \cite{Haw6, Haw7, Haw8}.
	
By extending the Hilbert space to include both positive- and negative-frequency photons, it is possible to localise a photon not only in space, but also in time.  In quantum physics, it has been a significant problem to define a time operator \cite{Alt}.  Whereas the position of a particle is associated with a position observable, time is only a parameter of the system.  Since in our quantisation all wave packets propagate causally, an operator can be defined that determines the time at which a particle will reach a particular position.  In investigations into a ``quantum relativity", where both space and time are placed on an equal quantum footing, a key concept is that of a closed and stationary universe \cite{Pag}.  The dynamics of these systems are constrained by a Wheeler-deWitt equation.  As the dynamical constraint for blips (\ref{dyn2}) is of this form, and defines a stationary state in the global Hilbert space of excitations in space-time, the blip quantisation may provide a useful and insightful scheme for the modelling of quantum clocks and the study of a quantum time.
	
One of the most significant features of this quantisation is the non-locality between the blip operators and the field observables.  This result is a direct consequence of the frequency-dependence of energy carried by the EM field.  In many schemes, localisation of a photon is synonymous with localisation of the field observables.  To fix the disparity between a probabilistic wave function and an energy carrying field, however, non-local or non-hermitian inner products are introduced.  Although often elegant, such inner products can be cumbersome, and may depend upon the boundary conditions of the system.  In the view of this quantisation, however, an excitation of the field is only localised if it satisfies the orthogonality relation (\ref{comm2}), and such non-local contributions must therefore exist.  Whilst this viewpoint may raise questions regarding instantaneous contributions to the field observables, the non-locality of the field observable is an important feature of the EM field.  For instance, in Ref.~\cite{Cas}, it is shown that these non-local field contributions are responsible for the Casimir effect between two parallel conducting plates.
	
The Schr\"odinger equation for light gives an alternative perspective on the particle behaviour of the free EM field.  By returning to principles of wave-particle duality, the existing theory of the EM field has been shown to fall short with regard to the propagation of localised particles, and as a result new physics has been unearthed.  The new quantisation that has followed provides a major change in our perspective on the interactions of photons and the way we describe them.  For instance, using this quantisation one can construct a local interaction Hamiltonian for a double-sided semi-transparent mirror; something that was not previously possible \cite{Jak2}.  Moreover, an investigation of blips in a cavity leads to a new perspective on the origin of the Casimir effect \cite{Cas}.  In future this quantisation may contribute towards a fuller understanding of well-known quantum effects, and may also provide the tools necessary for studying new ones. 

\section{Acknowledgements}
The author acknowledges financial support from the UK Engineering and Physical Sciences Research Council EPSRC [grant number EP/W524372/1].

\end{document}